\documentclass [a4paper,fleqn, 12pt]{article}
\usepackage{graphicx}
\usepackage[small]{subfigure,epsfig}

\usepackage {amsmath} \usepackage{amssymb} \usepackage{cite}

\newcommand{\cn}

\begin{document}


\title
{Exact solutions of the generalized $K(m,m)$ equations}

\author
{Nikolay A. Kudryashov, \and Svetlana G. Prilipko }

\date{Department of Applied Mathematics, National Research Nuclear University
MEPHI, 31 Kashirskoe Shosse,
115409 Moscow, Russian Federation}




\maketitle

\begin{abstract}

Family of equations, which is the generalization of the $K(m,m)$ equation, is considered. Periodic wave solutions for the family of nonlinear equations are constructed.

\end{abstract}






\section{Introduction}

Seeking to understand the role of nonlinear dispersion in the formation of patterns in liquid drops in 1993 Rosenau and Hyman \cite{Rosenau93} introduced a family of fully nonlinear $K(m,n)$ equations and also presented solutions of the $K(2,2)$ equation to illustrate the remarkable behavior of these equations. The $K(m,n)$ equations have the property that for certain $m$ and $n$ their solitary wave solutions have compact support. That is, they vanish identically outside a finite core region. These properties have a wide application in the fields of Physics and Mathematics, such as Nonlinear Optics, Geophysics, Fluid Dynamics and others.
Later, this equation was studied by various scientists worldwide \cite{Rosenau94, Rosenau98, Rosenau00, Tian05, Zhu06, Zhu06a, Odibat07, Xu08, Biswas08a,  Lou09, Bruzon09, Bruzon10, Biswas10}.

In this paper we construct periodic wave solutions for the following family of nonlinear partial differential equations

\begin{equation}\label{MainEq}
\frac{\partial{u}}{\partial{t}}+\sum_{k=0}^{N}\alpha_k \frac{\partial^{2k+1}{u^m}}{\partial{x^{2k+1}}}=0, \qquad N\geq 1, \qquad m\neq 1, \qquad \alpha_k \neq 0.
\end{equation}

The equation \eqref{MainEq}  is of order $ 2N+1 $ and depends on $N+2$ parameters denoted by $\alpha_0,...,\alpha_N,m$. This family contains a number of well-known generalizations of partial differential equations which were considered before \cite{Kudr88, Kudr91, Kudr92, Fu, Kudr05a, Kudr05b, Kudr07a, Kudr08b, Qin, Kudr08a, Lu, Kudr09a, Kudr09b, Kudr09c, Kudr10a}.

This paper is organized as follows. In Section 2 we describe a method which enables one to construct periodic wave solutions for the concerned family of nonlinear partial differential equations.
In Sections 3-6 we give several specific examples for some meanings of N.

\section{Method applied}

Applying traveling-wave variable
\begin{equation}\label{Travel}
u(x,t)=y(z), \qquad  z=x-C_0 t
\end{equation}
to Eq.\eqref{MainEq} and integrating the results yield the following Nth-order equation

\begin{equation} \label{MainEq1}
\sum_{k=0}^{N}\alpha_k \frac{d ^{2k}{y^m}}{d {z^{2k}}}-C_0\,y=0, \qquad N\geq 1, \qquad m\neq 1, \qquad \alpha_k \neq 0.
\end{equation}

The constant of integration is set to be zero. Substituting $y(z)=F(z)^p$ into
\[\alpha_n\,\frac{d^{2N}y^m}{dz^{2N}}-C_0\,y=0\]
we have $p=\frac{2\,N}{m-1}$. Note that Eq.\eqref{MainEq1} is an autonomous equation, and we can substitute  $z$ to $(z-z_0)$. We will take this fact into account in final solution, but we omit this substitution in our calculations. We search solutions of Eq.\eqref{MainEq1} in the form
\begin{equation}\label{Solut}
y(z)=(A_N)^{\frac{1}{m-1}}\cos^{\frac{2N}{m-1}}{(B_N(z-z_0))}.
\end{equation}

There is a remarkable property of a function $cos (B_N\,z)$. First of all we have to show expansion terms of Eq.\eqref{MainEq1}.

In the case $k=1$ we have the following expression
\begin{equation}
\begin{gathered}\label{der1}
\frac{d^2}{d z^2} \cos ^{{\frac {2m}{m-1}}}\left( B_1 \, z
 \right)  =-\frac{(2m{B_1})^{2}}{\left( m-1 \right)^2}  \cos^{{\frac {2m}{m-1}}}  \left( B_1  z \right)   + \\ \frac{2\,m
 \left( m+1 \right) {B_1}^{2}}{\left( m-1\right) ^{2}}\cos  ^{\frac{2}{m-1}}\left( B_1  z  \right).
\end{gathered}
\end{equation}

In the case $k=2$ we obtain
\begin{equation}
\begin{gathered}\label{der2}
\frac{d^4}{d z^4} \cos ^{{\frac {4m}{m-1}}}\left( B_2\, z
 \right)  =\frac{(4mB_2)^4}{(m-1)^4}\cos^{\frac{4m}{m-1}} \left( B_2 \,z\right) - \\ \frac{16mB_2^4(15m^3+11m^2+5m+1)}{(m-1)^4}\cos^{\frac{2(m+1)}{m-1}} \left( B_2 \,z\right) + \\ \frac{8mB_2^4(3m+1)(m+3)(m+1)}{(m-1)^4}\cos^{\frac{4}{m-1}} \left( B_2 \,z\right).
 \end{gathered}
\end{equation}

In the case $k=3$ we get
\begin{equation}
\begin{gathered}\label{der3}
\frac{d^6}{d z^6} \cos^{{\frac {4m}{m-1}}} \left( B_3\, z
 \right) =-\frac{{(6mB_3)}^{6}}{(m-1)^6}  \cos^{{\frac {6m}{m-1}}} \left( B_3\,z\right)+\\ \frac{96\,m{B_3}^{6} \left( 5\,m+1 \right)  \left( 7\,{m}^{2}
+m+1 \right) \left( 19\,{m}^{2}+7\,m+1 \right)}{(m-1)^6}  \cos^{{\frac {2(2\,m+1)}{m-1}}} \left( B_3\,z\right)- \\ -\frac{144\,m{B_3}^{6} \left( 2\,m+1 \right)\left( 5\,m+1 \right)
 \left( m+1 \right)  \left( 14\,{m}
^{2}+8\,m+5 \right) }{(m-1)^6}  \cos ^{{\frac {2(m+2)}{m-1}}}\left( B_3\,z\right)+ \\+ \frac{72\,m{B_3}^{6} \left( 5\,m+1 \right)  \left( 2\,m+1
 \right)  \left( m+5 \right)  \left( m+2 \right)  \left( m+1 \right)}{(m-1)^6}
\cos^{\frac{6}{m-1}} \left( B_3\,z \right).
 \end{gathered}
\end{equation}

In the general case $k=N$ derivative takes the form
\begin{equation}
\begin{gathered}\label{derN}
\frac{d^{2N}}{d z^{2N}}\cos ^{{\frac {2Nm}{m-1}}}\left( B_N \, z
 \right)  =(-1)^N \frac{(2NmB_N)^{2N}}{(m-1)^{2N}}\cos ^{\frac{2Nm}{m-1}}\left( B_N \,z\right) +\\
 +(-1)^{N+1}  \frac{B_N^{2N}M^{2N}_1}{(m-1)^{2N}}\cos^{\frac{2(Nm-m+1)}{m-1}} \left( B_N \,z\right) + (-1)^N \frac{B_N^{2N}M^{2N}_2}{(m-1)^{2N}}\cos^{\frac{2(Nm-2m+2)}{m-1}} \left( B_N \,z\right) +\\(-1)^{N+1} \frac{B_N^{2N}M^{2N}_3}{(m-1)^{2N}}\cos ^{\frac{2(Nm-3m+3)}{m-1}}\left( B_N \, z\right)  +... \\ -\frac{B_N^{2N}M^{2N}_{N-1}}{(m-1)^{2N}}\cos^{\frac{2(Nm-(N-1)m+N-1)}{m-1}} \left( B_N\, z\right)  + \frac{B_N^{2N}M^{2N}_N}{(m-1)^{2N}}\cos^{\frac{2N}{m-1}} \left( B_N\,z\right),
 \end{gathered}
\end{equation}
where $M^{2N}_1,...,M^{2N}_N$ are polynomials of $2N$ power.

Substituting Eq.\eqref{derN} into Eq.\eqref{MainEq1} we obtain the expression
\begin{equation}
\begin{gathered} \label{MainEq2}
A_N\left(\alpha_0-\frac{(2NmB_N)^2}{(m-1)^2}\alpha_1+\frac{(2NmB_N)^4}{(m-1)^4}\alpha_2-...+(-1)^N \frac{(2NmB_N)^{2N}}{(m-1)^{2N}}\alpha_N\right) \cos^{{\frac {2Nm}{m-1}}} \left( B_N\,z \right)+ \\
+A_N\left(\frac{B_N^2M_1^2}{(m-1)^2}\alpha_1-\frac{B_N^4M_1^4}{(m-1)^4}\alpha_2+...+(-1)^{N+1} \frac{B_N^{2N}M_1^{2N}}{(m-1)^{2N}}\alpha_N\right) \cos^{{\frac {2(Nm-m+1)}{m-1}}}
\left( B_N \, z\right)   + \\
+A_N\left(\frac{B_N^4M_2^2}{(m-1)^4}\alpha_2-\frac{B_N^6M_2^6}{(m-1)^6}\alpha_3+... +(-1)^N \frac{B_N^{2N}M_2^{2N}}{(m-1)^{2N}}\alpha_N\right) \cos^{{\frac {2(Nm-2m+2)}{m-1}}} \left( B_N\,z\right)+ ...+\\
+A_N\left( \frac{B_N^{2(N-1)}M_{N-1}^{2(N-1)}}{(m-1)^{2(N-1)}}\alpha_{N-1}-\frac{B_N^{2N}M_{N-1}^{2N}}
{(m-1)^{2N}}\alpha_N\right) \cos^{{\frac {2(Nm-(N-1)m+N-1)}{m-1}}} \left( B_N\,z\right)+ \\
+\left(\frac{B_N^{2N}M_{N}^{2N}}{(m-1)^{2N}}\alpha_N A_N-C_0\right) \cos^{{\frac {2N}{m-1}}} \left( B_N \,z\right)=0.
\end{gathered}
\end{equation}

Equating coefficients at powers of $\cos \left( B_N\,z\right)$ to zero yields an algebraic system. Solving this system we obtain the values of parameters $A_N, B_N$ and correlations on the coefficients $\alpha_{0},...,\alpha_{N}$.

\section{Periodic wave solutions of the K(m,m) equation}

The first member of the family \eqref{MainEq} in the case $N=1$ takes the form
\begin{equation}
\begin{gathered} \label{Main1}
u_{t}+\alpha_0(u^{m})_{x}+\alpha_1(u^{m})_{3,x}=0,\quad (u^{m})_{3,x}=\frac{d^3u^m}{dx^3},\\  m\neq 1,\quad \alpha_0,\alpha_1 \neq 0.
\end{gathered}
\end{equation}

Taking the traveling wave ansatz \eqref{Travel} into account, we have the equation with respect to $y(z)$
\begin{equation} \label{Main11}
\alpha_1(y^{m})_{2,z}+\alpha_0 y^{m}-C_{0}\,y=0, \quad (y^{m})_{2,z}=\frac{d^2y^m}{dz^2}.
\end{equation}

Following the procedure suggested in the previous section we obtain the equation
\begin{equation}
\begin{gathered}\label{Main12}
\left( \alpha_0-{\frac {{(2mB_1)}^{2}}{
 \left( m-1 \right) ^{2}}}\alpha_1 \right) {\it A_1}\,  \cos^{{\frac {2m}{m-1}}}
 \left( B_1\,z \right) + \\ +\left( \frac{2\,m\left( m+1 \right) }{(m-1)^2}{B_1}^{2}\,{\it A_1}\,\alpha_1-C_0\right) \cos ^{\frac{2}{m-1}} \left( B_1\,z\right) = 0.
\end{gathered}
\end{equation}

Equating coefficients at powers of $\cos \left( B_1\, z\right)$ to zero we get values of parameters $A_1, B_1$
\begin{equation}\label{Main1A}
A_1=\frac{2C_0 m}{\alpha_0(m+1)},\qquad m\neq -1,
\end{equation}

\begin{equation}\label{Main1B}
B_1=\sqrt{\frac{\alpha_0}{\alpha_1}}\frac{(m-1)}{2m},\qquad m\neq 0.
\end{equation}

Solutions of Eq.\eqref{Main12} are the following
\begin{equation}\label{Main1y}
y(z)=\left(\frac{2C_0 m}{\alpha_0(m+1)}\right)^\frac{1}{m-1}\cos^\frac{2}{m-1} \left( \sqrt{\frac{\alpha_0}{\alpha_1}}\frac{(m-1)}{2m} \left( z-{\it z_0} \right)\right).
\end{equation}

In the case $\alpha_0=1,\alpha_1=1$ formula \eqref{Main1y} is a solution of a well-known $K(m,n)$ equation with $n=m$ \cite{Rosenau93}. 

\section{Periodic wave solutions of Eq.\eqref{MainEq1} in the case N=2}

In this section we will construct exact solutions of the family \eqref{MainEq} at $N=2$. In this case  Eq. \eqref{MainEq1} takes the form
\begin{equation}
\begin{gathered} \label{Main21}
u_{t}+\alpha_0(u^{m})_{x}+\alpha_1(u^{m})_{3,x}+\alpha_2(u^{m})_{5,x}=0,\quad (u^{m})_{5,x}=\frac{d^5u^m}{dx^5},\\ m\neq 1,\quad \alpha_0,\alpha_1,\alpha_2\neq 0.
\end{gathered}
\end{equation}

Making the substitution \eqref{Travel} into \eqref{Main21} and integrating the result we obtain
\begin{equation}\label{Main22}
 \alpha_2 (y^{m})_{4,z}+\alpha_1 (y^{m})_{2,z}-C_{0}y+\alpha_0 y^{m}=0,\quad (y^{m})_{4,z}=\frac{d^4y^m}{dz^4}.
\end{equation}

Substitution of  solution \eqref{Solut} into \eqref{Main22} allows us to get the following equation
\begin{equation}
\begin{gathered}\label{Main23}
A_2\left(\alpha_0-\frac{(4mB_2)^2}{(m-1)^2}\alpha_1+\frac{(4mB_2)^4}{(m-1)^4}\alpha_2\right)  \cos^{{\frac {4m}{m-1}}} \left( B_2\, z\right)   + \\ + A_2\left(\frac{4m(3m+1)}{(m-1)^2}B_2^2\alpha_1-\frac{16m(15m^3+11m^2+5m+1)}{(m-1)^4}B_2^4\alpha_2\right) \cos^{{\frac {2(m+1)}{m-1}}} \left( B_2\,z\right) + \\ +\left(\frac{8m(3m^3+13m^2+13m+3)}{(m-1)^4}B_2^4A_2\alpha_2-C_0\right) \cos ^{{\frac {4}{m-1}}} \left( B_2\, z\right)  =0.
\end{gathered}
\end{equation}

Solving this equation we obtain
\begin{equation}\label{Main2A}
A_2=\frac{2^3 C_0m(m+1)}{\alpha_0(3m+1)(m+3)},\qquad m\neq -\frac{1}{3},3,
\end{equation}

\begin{equation}\label{Main2B}
B_2=\sqrt{\frac{\alpha_0}{\alpha_1}}\frac{(m-1)\sqrt{5m^2+2m+1}}{4m(m+1)},\qquad m\neq 0,-1,
\end{equation}

\begin{equation}\label{Main2a}
\alpha_2=\frac{4(m+1)^2 \alpha_1^2}{m(5m^2+2m+1)^2\alpha_0 }.
\end{equation}

Formula \eqref{Solut} in the case $N=2$ is the following
\begin{equation}\label{Main2y}
y(z)=\left(\frac{2^3 C_0m(m+1)}{\alpha_0(3m+1)(m+3)}\right)^\frac{1}{m-1}\cos^\frac{4}{m-1} \left( \sqrt{\frac{\alpha_0}{\alpha_1}}\frac{(m-1)\sqrt{5m^2+2m+1}}{4m(m+1)} \left( z-{\it z_0} \right)\right).
\end{equation}


\section{Periodic wave solutions of Eq.\eqref{MainEq1} in the case N=3}

Formula \eqref{MainEq} in the case $N=3$ is
\begin{equation}
\begin{gathered}\label{Main3}
u_{t}+\alpha_0(u^{m})_{x}+\alpha_1(u^{m})_{3,x}+\alpha_2(u^{m})_{5,x}+\alpha_3 (u^{m})_{7,x}=0,\\ \\ (u^{m})_{7,x}=\frac{d^7u^m}{dx^7}, \quad m\neq 1,\quad \alpha_0,\alpha_1,\alpha_2 ,\alpha_3\neq 0.
\end{gathered}
\end{equation}

Using traveling wave reduction \eqref{Travel}, the following ordinary differential equation takes the form
\begin{equation}\label{Main33}
\alpha_3 (y^{m})_{6,z}+\alpha_2 (y^{m})_{4,z} +\alpha_1(y^{m})_{2,z}+\alpha_0 y^{m}-C_{0}y=0, \quad (y^m)_{6,z}=\frac{d^6 y^m}{dz^6}.
\end{equation}

In the case $N=3$ from Eq.\eqref{MainEq2}  we obtain values of the parameters $A_3, B_3$
\begin{equation}\label{Main3A}
A_3=\frac{2^3 C_0(m+2)(2m+1)m}{\alpha_0(5m+1)(m+5)(m+1)},\quad m\neq -\frac{1}{5},-1,-5,
\end{equation}

\begin{equation}\label{Main3B}
B_3=\sqrt{\frac{\alpha_0}{\alpha_1}}\frac{(m-1)\sqrt{49m^4+92m^3+78m^2+20m+4}}{6(2m+1)(m+2)m},\qquad m\neq 0, -\frac{1}{2}, -2.
\end{equation}

Relations between the coefficients $\alpha_0$, $\alpha_1$, $\alpha_2$ and $\alpha_3$ are the following
\begin{equation}\label{Main32a}
\alpha_2=\frac{9(2m+1)^2(m+2)^2m^2(14m^2+8m+5)\alpha_1^2}{(49m^4+92m^3+78m^2+20m+4)^2\alpha_0},
\end{equation}

\begin{equation}\label{Main33a}
\alpha_3=\frac{81m^4(2m+1)^4(m+2)^4\alpha_1^3}{(49m^4+92m^3+78m^2+20m+4)^3\alpha_0^2 }.
\end{equation}

Solutions of Eq.\eqref{Main33} take the form
\begin{equation}\label{Main3y}
y(z)=(A_3)^{\frac{1}{m-1}}\cos^{\frac{6}{m-1}}{(B_3(z-z_0))},
\end{equation}
where values of parameters $A_3$, $B_3$ are given by formulae \eqref{Main3A} and \eqref{Main3B}.

In the case $m=3$, $C_0=1$, $z_0=0$, $\alpha_0=1$ and $\alpha_1=1$ solution of Eq. \eqref{Main3y} is presented in Fig. $1$.

\begin{figure}[h]
  \centering
    \includegraphics[width=0.55\textwidth]{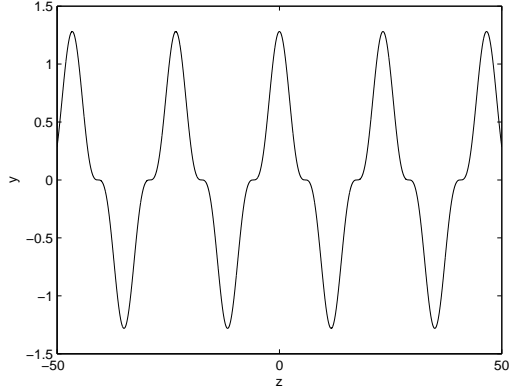}
    \caption{Solution of Eq. \eqref{Main3y} in the case $N=3$, $m=3$, $C_0=1$, $z_0=0$, $\alpha_0=1$, $\alpha_1=1$.}
  \label{F1}
\end{figure}

Note that in the case $N=C(2l+1)$, where C is an arbitrary constant and $l=1,2,..$, solutions \eqref{Solut} are alike solution given in Fig. $1$.

\section{Periodic wave solutions of Eq.\eqref{MainEq1} in the case N=4}

Let us look for exact solutions of the family \eqref{MainEq} in the case N=4
\begin{equation}
\begin{gathered}\label{Main4}
u_{t}+\alpha_0(u^{m})_{x}+\alpha_1(u^{m})_{3,x}+\alpha_2(u^{m})_{5,x}+\alpha_3 (u^{m})_{7,x}+\alpha_4(u^{m})_{9,x}=0,\\  (u^{m})_{9,x}=\frac{d^9u^m}{dx^9},  \quad m\neq 1,\quad \alpha_0,\alpha_1,\alpha_2 ,\alpha_3,\alpha_4\neq 0.
\end{gathered}
\end{equation}

This equation possesses the traveling wave reduction \eqref{Travel} with $y(z)$ satisfying the equation
\begin{equation}
\begin{gathered}\label{Main44}
\alpha_4 (y^{m})_{8,z}+\alpha_3(y^{m})_{6,z}+\alpha_2 (y^{m})_{4,z}+\alpha_1(y^{m})_{2,z}+\alpha_0 y^{m}-C_{0}y=0,\\ (y^m)_{8,z}=\frac{d^8 y^m}{dz^8}.
\end{gathered}
\end{equation}

Following the procedure suggested in section 2 we obtain values of the parameters $A_4, B_4$
\begin{equation}\label{Main4A}
A_4=\frac{2^7 C_0m(m+3)(3m+1)(m+1)}{\alpha_0(3m+5)(7m+1)(m+7)(5m+3)},  \qquad m\neq -\frac{1}{7},-\frac{5}{3},-\frac{3}{5},-7,
\end{equation}

\begin{equation}
\begin{gathered}\label{Main4B}
B_4=\sqrt{\frac{\alpha_0}{\alpha_1}}\frac{(m-1)\sqrt{205m^6+830m^5+1423m^4+1108m^3+443m^2+78m+9}} {8(3m+1)(m+3)(m+1)m}, \\ \qquad m\neq 0,-\frac{1}{3},-1,-7,
\end{gathered}
\end{equation}

and correlations on the coefficients
\begin{equation}\label{Main42a}
\alpha_2=\frac{4(m+3)^2(3m+1)^2(m+1)^2m^2(273m^4+508m^3+518m^2+188m+49) \alpha_1^2} {(205m^6+830m^5+1423m^4+1108m^3+443m^2+78m+9)^2\alpha_0},
\end{equation}

\begin{equation}\label{Main43a}
\alpha_3=\frac{128(15m^2+10m+7)(m+3)^4(3m+1)^4(m+1)^4m^4\alpha_1^3} {(205m^6+830m^5+1423m^4+1108m^3+443m^2+78m+9)^3\alpha_0^2},
\end{equation}

\begin{equation}\label{Main44a}
\alpha_4=\frac{1024(m+3)^6(3m+1)^6(m+1)^6m^6\alpha_1^4} {(205m^6+830m^5+1423m^4+1108m^3+443m^2+78m+9)^4\alpha_0^3}.
\end{equation}

Eq.\eqref{Solut} takes the form
\begin{equation}\label{Main4y}
y(z)=(A_4)^{\frac{1}{m-1}}\cos^{\frac{8}{m-1}}{(B_4(z-z_0))},
\end{equation}
where $A_4$ and $B_4$ are determined by formulae \eqref{Main4A} and \eqref{Main4B}.

In the case $m=2$ solution \eqref{Main4y} is given in Fig.$2$.
\begin{figure}[h]
  \centering
    \includegraphics[width=0.55\textwidth]{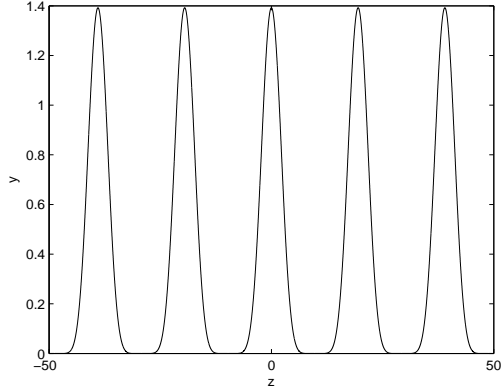}
    \caption{Solution of Eq. \eqref{Main4y} in the case $N=4$, $m=2$, $C_0=1$, $z_0=0$, $\alpha_0=1$, $\alpha_1=1$.}
  \label{F2}
\end{figure}

Note, that the amplitude and period of the traveling wave solution are growing with the increasing of $N$. The dependence of amplitude on $N$ is illustrated in Fig. $3$ in the case $m=2$.

\begin{figure}[h]
  \centering
    \includegraphics[width=0.55\textwidth]{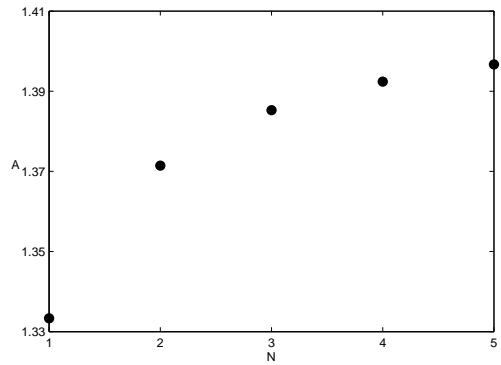}
    \caption{The dependence of amplitude of the traveling wave solution on $N$.}
  \label{}
\end{figure}



In the general case solution of the family of equations \eqref{MainEq} takes the form \eqref{Solut}, where parameters $A_N, B_N$ can be written as

\begin{equation}\label{MainNA}
A_N=\frac{2^N C_0\underset{j=0}{\overset{N-1}\prod}((N-j)m+j)}{\alpha_0\underset{j=0}{\overset{N-1}\prod}
((2(N-j)-1)m+2j+1)},
\end{equation}

\begin{equation}\label{MainNB}
B_N=\sqrt{\frac{\alpha_0}{\alpha_1}}\frac{(m-1)P^{N-1}_N}{\underset{j=0}{\overset{N-1}\prod}((N-j)m+j)},
\end{equation}
for $m\neq -\frac{2j+1}{2(N-j)-1}$, $-\frac{j}{N-j}$ $(j=0,1,...,N-1)$.
$P^{N-1}_N$  are polynomials of $N-1$ power.
Relations for values of the coefficients $\alpha_k$ $(k=0,1,..,N)$ are found from \eqref{MainEq2}.

\section{Conclusion}

Let us formulate shortly the results of this paper. We have studied the generalized $K(m,m)$ equations. Taking into consideration the traveling wave ansatz we have found the periodic wave solutions for a family of nonlinear partial differential equation \eqref{MainEq}. This family generalizes a well-known K(m,n) equation in the case $n=m$. Formula for the amplitude of the traveling wave in the general case is given. Exact solutions for the cases $N=1,2,3,4$ of the family \eqref{MainEq} are presented.

\section*{Acknowledgment}

This work was supported by the Federal target programm "Research and scientific-pedagogical personnel of innovation in Russia" of 2009-2011.

\end{document}